\documentclass[10pt,aps,superscriptaddress,physrev,twocolumn]{revtex4-2}

\usepackage{amsmath}
\usepackage{amssymb}
\usepackage{hyperref}
\usepackage{appendix}

\usepackage[dvipsnames]{xcolor} 
\usepackage{soul}

\begin{document}

\title{Quantum Gravity, de Sitter Space, and Normalizability}

\author{Stephon Alexander}
\affiliation{Brown Theoretical Physics Center, Department of Physics, Brown University, Providence, RI 02912, USA}
\author{Heliudson Bernardo}
\affiliation{Department of Physics \& Astronomy, University of Lethbridge, Lethbridge, AB T1K 3M4, Canada}
\author{Jacob Kuntzleman}
\author{Max Pezzelle}
\affiliation{Brown Theoretical Physics Center, Department of Physics, Brown University, Providence, RI 02912, USA}

\newcommand{\HB}[1]{{\color{blue}#1}}

\newcommand{\JK}[1]{{\color{purple}#1}}

\newcommand{\red}[1]{{\color{red}#1}}

\begin{abstract}

We propose a resolution to the longstanding problem of perturbative normalizability in canonical quantum gravity of the Lorentzian Chern-Simons-Kodama (CSK) state with a positive cosmological constant in four dimensions. While the CSK state is an exact solution to the Hamiltonian constraint in the self-dual formulation and semiclassically describes de Sitter spacetime, its physical viability has been questioned due to apparent nonnormalizability and CPT asymmetry.  Starting from a nonperturbative holomorphic inner product derived from the reality conditions of the self-dual Ashtekar variables, we show that the linearization, in terms of gravitons, of the CSK state is perturbatively normalizable for super-Planckian cosmological constant. Furthermore, we demonstrate that a rotation in phase space, a generalization of Thiemann’s complexifier, can render the full perturbative state normalizable for all $\Lambda$ by analytically continuing the non-convergent modes in phase space. This provides the first concrete realization of a CPT-breaking, yet normalizable, gravitational vacuum state rooted in a nonperturbative quantum gravity framework. Our results establish the CSK state—long thought formal—as a viable candidate for the ground state of quantum gravity in de Sitter space.
\end{abstract}

\maketitle

\section{Introduction}

A foundational goal of canonical quantum gravity is the identification of a physically meaningful vacuum state: one that solves the quantum Hamiltonian, diffeomorphism, and gauge constraints, admits a sensible semiclassical limit, and supports a perturbative description of gravitons. In the Ashtekar formulation of general relativity with a positive cosmological constant \cite{Ashtekar:1986yd, Ashtekar:1987gu}, the Chern-Simons-Kodama (CSK) state~\cite{Kodama:1990sc} is an exact and elegant solution to the quantum constraints. It is sharply peaked on de Sitter spacetime and encapsulates the self-dual sector of gravity through its Chern--Simons form. Yet, despite its appeal, the CSK state has long been plagued by concerns over its nonnormalizability, its complex structure, and its apparent violation of CPT symmetry~\cite{Witten:2003mb}.

The key difficulty has been that under the naïve inner product --- the formal analog of the Yang--Mills measure \cite{Freidel:2003pu} --- the Kodama state is exponentially divergent~\cite{Smolin:2002sz}. However, this conclusion hinges critically on the choice of inner product. In recent work~\cite{Alexander:2022ocp}, one of us and others have shown that a consistent quantization of self-dual gravity must employ a holomorphic inner product, one derived from enforcing the quantum analogs of the classical reality conditions. When this physically motivated inner product is adopted, a new perspective on the CSK state arises.

In this work, we analyze linearized fluctuations around de Sitter spacetime using the holomorphic inner product introduced in~\cite{Alexander:2022ocp}, tailored to the complex Ashtekar connection. For super-Planckian values of the cosmological constant $\Lambda$, we show that the CSK state is perturbatively normalizable while, for smaller values, it factorizes into normalizable and nonnormalizable sectors. The super-Planckian-$\Lambda$ result establishes a normalizable graviton vacuum in a background de Sitter geometry~\cite{Magueijo:2010ba}.

Furthermore, we propose that a generalized, mode-by-mode Wick rotation---introduced by Thiemann---can analytically continue only the divergent modes problematic at sub-Planckian values of the cosmological constant, converting the full state into a perturbatively normalizable one~\cite{Thiemann:1995ug, Ashtekar:1995qw}, rendering the norm finite.

Our results offer the first perturbatively well defined realization of a gravitational vacuum wavefunctional rooted in the nonperturbative canonical quantization of general relativity. We argue that this construction opens a viable path forward for quantum cosmology in the Ashtekar framework and invites a reinterpretation of the Chern-Simons-Kodama state --- not as a mathematical curiosity, but as a physical state of the quantum universe.

In Sec. \ref{sec:quantization}, we outline the canonical quantization of general relativity in the self-dual formalism, emphasizing in subsection \ref{subsec:holomorphic_representation} the many subtleties that arise in interpreting normalizability of wavefunctionals in the self-dual connection representation. In Sec. \ref{sec:kodama}, we review the Kodama state and discuss the question of its normalizability, both perturbatively and nonperturbatively. In Sec. \ref{sec:linearization}, we present the main calculation of the paper, in which we linearize the nonperturbative inner product (\ref{eq:4d_inner_product}) about de Sitter space and demonstrate that it renders the perturbative Kodama state normalizable for sufficiently large $\Lambda$. Finally, in Sec. \ref{sec:discussion}, we discuss various interpretations of this result, most importantly how an extension of the phase-space Wick rotation of Thiemann could render the CSK state normalizable for all values of $\Lambda$.

\section{Canonical Quantization}
\label{sec:quantization}

In this section we outline the canonical quantization of general relativity in its self-dual formulation. We introduce Ashtekar's variables, placing particular emphasis on the non-orthonormal eigenstates of the Ashtekar connection operator.
In the final subsection, we argue that the na\"{i}ve inner product is inconsistent with the commutation relations of the quantum theory, forcing one to seek alternatives. This motivates the discussion in Sec. \ref{sec:kodama}, where we see that (\ref{eq:4d_inner_product}) is a viable inner product.

Our index conventions are the following. Greek letters $\{\alpha,\beta,...\}$ denote world indices and capital Latin letters $\{I,J,..\}$ denote frame indices, both of which run from 0 to 3. Lowercase Latin letters from the beginning of the alphabet $\{a,b,...\}$ denote spatial world indices and those from the middle of the alphabet $\{i,j,...\}$ denote spatial frame indices, both of which run from 1 to 3. Additionally, we work in the $\{-,+,+,+\}$ metric signature.

\subsection{Self-dual Formulation of General Relativity}
\label{subsec:self-dual}

Einstein's field equations may be obtained from the action
\begin{equation} 
    S_{\rm EC} =\frac{1}{32\pi G} \int\epsilon_{IJKL} \, e^I\wedge e^J\wedge\bigg[ R^{KL} - \frac{\Lambda}{6} e^K\wedge e^L\bigg]
    \label{eq:gr_action}
\end{equation}
where $e^I = e^{I}_{\alpha} \mathrm{d}x^\alpha$ is the metric \textit{vierbein}, $\Lambda$ is the cosmological constant, and we have introduced the curvature two-form $R^{IJ}$ of the \textit{spin connection} $\omega_{IJ}$,
\begin{equation}
    R^{IJ} = \mathrm{d}\omega^{IJ} + \omega^I_{\mathrm{\ }K}\wedge \omega^{KJ}.
\end{equation}
Upon varying with respect to $\omega$, we find
\begin{equation}
    D(\epsilon_{IJKL}e^K\wedge e^L) = 0
    \label{eq:omega_eom}
\end{equation}
where, for a mixed-rank group tensor $U^I_{\ J}$, the exterior covariant derivative acts as $DU^I_{\ J} = \mathrm{d}U^I_{\ J} + \omega^I_{\ K}\wedge U^K_{\ J} - \omega^K_{\ J}\wedge U^I_{\ K}$. For non-degenerate vierbein, this equation implies that the \textit{torsion} $T^I = \mathrm{d}e^I + \omega^I_{\mathrm{\ }J}\wedge e^J$ is zero, and further implies that the spin connection is fully determined by the frame field and its derivatives. The equation of motion for the frame field is
\begin{equation}
    \epsilon_{IJKL}e^J\wedge \bigg(R^{KL} - \frac{\Lambda}{3}e^K\wedge e^L\bigg) = 0
\end{equation}
which are the Einstein field equations. Indeed, when $T^I = 0$, $S_{\rm EC}$ is proportional to the Einstein-Hilbert action,
\begin{equation}
    S_{\rm EH} = \frac{1}{16\pi G} \int d^4 x \, \sqrt{-g} (R - 2\Lambda).
\end{equation}

The field equations can also be derived using self-dual variables. This is accomplished after complexifying the local group SO$(1,3)$ and working only with variables which are self-dual with respect to the group indices $\{I, J,\dots\}$ \cite{Jacobson:1987yw,Jacobson:1988yy,Samuel:1987td}. In terms of quantities valued in the sl$(2, \mathbb{C})^{+}$, the complex-valued action is (see e.g. \cite{Celada:2016jdt,Sahlmann:2023eqt})
\begin{equation}
    S_{\rm SD} = \frac{i}{16\pi G}\int\bigg[\Sigma_i\wedge F^i - \frac{\Lambda}{6}\Sigma_i \wedge \Sigma^i\bigg],
    \label{eq:sd_action_compressed}
\end{equation}
where
\begin{equation}
    \Sigma^i = ie^0 \wedge e^i -\frac{1}{2}\epsilon^{ijk}e_j \wedge e_k
\end{equation}
is the self-dual projection of the two-form $ie^0 \wedge e^i$ and
\begin{equation}
    F^i = \mathrm{d}A^i + \frac{1}{2}\epsilon^{ijk}A_j\wedge A_k,
\end{equation}
is the curvature of the \textit{self-dual connection}, 
\begin{equation}
    A^i = i\omega^{0i} - \frac{1}{2}\epsilon^{ijk}\omega_{jk},
    \label{eq:ashtekar_connection}
\end{equation}
which is valued in the sl$(2, \mathbb{C})^{+}$ gauge algebra. For vanishing torsion, this is the Ashtekar connection \cite{Ashtekar:1986yd,Ashtekar:1987gu}.

The complexification of the local group introduces no modification to the classical dynamics of $S_{\rm EC}$. In fact, in terms of real variables, 
\begin{equation}
    S_{\rm SD} = S_{\rm EC}[e, \omega] + \frac{i}{16\pi G} \int e_I\wedge e_J \wedge R^{IJ}.
    \label{eq:sd_action_expanded}
\end{equation}
From the first term we recover ordinary Einstein-Cartan theory; the extra term is known as the \textit{Holst term} \cite{Holst:1995pc}. It can be written in terms of the torsion:
\begin{equation}
    e^I \wedge e^J \wedge R_{IJ} = e^I \wedge DT_I.
\end{equation}
The Holst term is neither a total derivative, nor is it identically zero. For $T^I = 0$, however, it manifestly vanishes. Varying $S_{\rm SD}$ with respect to $\omega$ sets $T^I \equiv 0$, such that $S_{\rm SD}$ is classically equivalent to both $S_{\rm EC}$ and $S_{\rm EH}$. They are not equivalent quantum-mechanically, however, as we now discuss.

The action $S_{\rm SD}$ is obtained from $S_{\rm EH}$ by adding dynamically-vanishing extra structure, in this case the torsion $T^I$\footnote{Of course, there is an infinity of actions, all classically equivalent to $S_{\rm EH}$, which are obtained in a similar way.}. When quantizing general relativity with the Einstein-Hilbert action $S_{\rm EH}$, one ignores quantum fluctuations in the torsion, treating vanishing of torsion as an operator identity, while when quantizing $S_{\rm SD}$ quantum torsion fluctuations enter naturally into the analysis. Practically, this means that the path integral defined by $S_{\rm SD}$ includes summations over connections with nonvanishing torsion, while all paths that appear in the $S_{\rm EH}$ path integral have $T^I \equiv 0$ by fiat. This will become relevant for our considerations.

We stress that the omission of torsion fluctuations from the $S_{\rm EH}$ path integral does not by itself disqualify the Einstein-Hilbert quantum theory relative to the self-dual theory. The ultimate arbiter of the correct quantum action is nature; if quantum torsion fluctuations are detected, then the $S_{\rm EH}$ action will be falsified. However, there are good reasons (besides the algebraic specialness of $S_{\rm SD}$) to expect the correct theory to have nonvanishing torsion. Indeed, not only is $S_{\rm SD}$ consistent with the symmetries of general relativity, the inclusion of torsion in the theory is also necessary to allow coupling between fermions and gravity \cite{Jacobson:1988qta}. For these reasons, in this work we take the self-dual action (\ref{eq:sd_action_compressed}) as our starting point.

\subsection{Hamiltonian Formulation}
\label{subsec:hamiltonian}

Here, we show that the self-dual variables $\Sigma^i$ and $A^i$ arise naturally in the Hamiltonian formulation of general relativity, hence their importance in canonical quantum gravity. To construct the Hamiltonian formulation, we first need to foliate the spacetime with a family of non-intersecting spacelike hypersurfaces \cite{PhysRev.160.1113,Misner:1973prb, Poisson:2009pwt}. Let $t$ be a scalar field on the spacetime such that $t = \mathrm{constant}$ specifies a hypersurface $\Sigma_t$. The gradient $n_\alpha = \partial_\alpha t$ is a future-directed timelike vector. We shall install coordinates $\{y^a\}$ on the hypersurfaces $\Sigma_t$ using a geodesic congruence $\gamma$. The tangent vectors on $\Sigma_t$ are given by
\begin{equation}
    \hat{e}^\alpha_a = \bigg(\frac{\partial x^\alpha}{\partial y^a}\bigg)\Bigg|_t.
\end{equation}
The tangent vector $t^\alpha$ to the congruence satisfies $t^\alpha\partial_\alpha t = 1$. It can be decomposed into one part parallel to the hypersurface $\Sigma_t$ and one part orthogonal to it:
\begin{equation}
    t^\alpha = Nn^\alpha + N^a \hat{e}^\alpha_a
\end{equation}
where $N$ is the lapse and $N^a$ is the shift vector. In terms of this 3+1 decomposition, the metric becomes
\begin{equation}
    \mathrm{d}s^2 = -N^2\mathrm{d}t^2 + h_{ab}(N^a \mathrm{d}t + \mathrm{d}y^a)(N^b \mathrm{d}t + \mathrm{d}y^b)
\end{equation}
where $h_{ab} = g_{\alpha\beta}\hat{e}^\alpha_a \hat{e}^\beta_b$. The \emph{densitized triad} is defined as
\begin{equation}
    E^a_i = -\sqrt{h}\epsilon^a_{\mathrm{\ }bc}\Sigma_i^{\mathrm{\ }bc}
    \label{eq:densitized_triad}
\end{equation}
where $\Sigma_i^{\mathrm{\ }bc}$ is the self-dual two-form associated with the triad.  It can be shown that the action for general relativity in its self-dual form is \cite{Celada:2016jdt,Sahlmann:2023eqt}
\begin{multline}
    S_{\rm SD} = \frac{1}{8\pi  Gi}\int d^4x \bigg[E^a_i\mathcal{L}_t A^i_a \\ + \frac{iN}{2\sqrt{h}}\epsilon^{ijk}E^a_i E^b_j\bigg(F_{kab} + \frac{\Lambda}{3}\epsilon_{abc}E^c_k\bigg)\\ + (t^\alpha A^i_\alpha)\mathcal{D}_a E^a_i + N^a E^b_i F^i_{ba}\bigg].
    \label{eq:constraint_action}
\end{multline}
Notice that the time derivatives of $N$, $N^a$, and $t^\alpha A^i_\alpha$ do not appear in the action. This means that the equations of motion of these fields are constraints on the system. The second term corresponds to the \textit{Hamiltonian constraint}, the third to the \textit{Gauss constraint}, and the fourth to the \textit{diffeomorphism constraint}. The Gauss constraint enforces invariance under small gauge transformations, while the diffeomorphism constraint enforces invariance under spatial diffeomorphisms. If we enforce that these constraints are satisfied from the outset, then the space of solutions is reduced to \textit{superspace}. What is left to enforce is the Hamiltonian constraint,
\begin{equation}
    \mathcal{H} = -\frac{1}{16\pi G}\epsilon^{ijk}E^a_i E^b_j\bigg(F_{kab} + \frac{\Lambda}{3}\epsilon_{abc}E^c_k\bigg) =0
    \label{eq:hamiltonian_constaint}
\end{equation}
where we have scaled out the factor of $\sqrt{h}$ from the denominator so that the constraint is polynomial in the fields $E^a_i$ and $A^i_a$.

\subsection{The Connection Representation is a Holomorphic Representation}
\label{subsec:holomorphic_representation}

Having obtained the Hamiltonian constraint, we may proceed to quantization of the self-dual theory. From the self-dual action, we find that the densitized triad and connection satisfy the canonical Poisson bracket
\begin{equation}
    \{A^i_a(\vec{x}),E^b_j(\vec{y})\} = 8\pi Gi \delta^b_a \delta^i_j \delta^{(3)}(\vec{x}-\vec{y}).
    \label{eq:poisson}
\end{equation}
Upon quantization, then, the connection and triad operators satisfy the canonical commutation relation
\begin{equation}
    [\hat{A}^i_a(\vec{x}),\hat{E}^b_j(\vec{y})] = - \ell_{\rm Pl}^2 \delta^b_a \delta^i_j \delta^{(3)}(\vec{x}-\vec{y}).
    \label{eq:self-dual_commutation_relations}
\end{equation}
where $\ell_{\rm Pl}^2 = 8\pi G\hbar$ is the square of the Planck length. Since $\hat{E}^a_i(\vec{x})$ is self-adjoint, $\hat{E}^a_i(\vec{x}) = \hat{E}^{a\dagger}_i(\vec{x})$, and since
\begin{equation}
    [\hat{E}^a_i(\vec{x}), \hat{E}^b_j(\vec{y})] = 0
\end{equation}
for all $i, j$; $a, b$; and $\vec{x}, \vec{y}$, the eigenstates $\rvert E \rangle$ of the $\hat{E}^a_i(\vec{x})$ operator furnish an orthonormal basis of the Hilbert space $\mathcal{H}$ of the quantum theory. With these states, we construct wavefunctionals $\Psi[E] \equiv \langle E \rvert \Psi \rangle$ from arbitrary states $\rvert \Psi \rangle \in \mathcal{H}$. In this \textit{triad representation}, we have
\begin{equation}
\begin{split}
    \hat{E}^a_i(\vec{x}) \Psi[E] &= E^a_i(\vec{x}) \Psi[E],
    \\
    \hat{A}^i_a(\vec{x}) \Psi[E] &= -\ell_{\rm Pl}^2 \frac{\delta}{\delta E^a_i(\vec{x})} \Psi[E],
\end{split}
\end{equation}
where the action of $\hat{A}^i_a(\vec{x})$ follows from the commutation relations (\ref{eq:self-dual_commutation_relations}).

Since $[\hat{A}^a_i(\vec{x}), \hat{A}^b_j(\vec{y})] = 0$, we can similarly construct a basis of eingenstates $\rvert A \rangle$ of the connection operator $\hat{A}^i_a(\vec{x})$ and the corresponding \textit{connection-representation} wavefunctionals $\Psi[A] \equiv \langle A \rvert \Psi \rangle$. The construction is hindered, however, by the fact that the connection operator is not self-adjoint. Indeed, from the definition (\ref{eq:ashtekar_connection}), we find
\begin{equation}
    [\hat{A}^i_a(\vec{x}), \hat{A}^{j\dagger}_b(\vec{y})] = \ell_{\rm Pl}^2 \epsilon^{jkl} \frac{\delta}{\delta E^a_i(\vec{x})}\omega_{bkl}[E] \neq 0.
\end{equation}
Consequently, the $| A \rangle$ basis need not be orthonormal. In fact, the eigenstates of $\hat{A}^i_a(\vec{x})$ furnish an \textit{overcomplete basis}, complicating the interpretation of wavefunctionals like $\Psi[A]$ \cite{Kodama:1990sc}. Importantly, it is possible for normalizable wavefunctionals, when written in terms of an overcomplete basis, to superficially appear nonnormalizable. To better understand this fact, let us briefly review a simple analogue from single-particle quantum mechanics.

\subsubsection{Holomorphic Inner Product for the Harmonic Oscillator}

Consider a quantum harmonic oscillator with $m = \omega = 1$ and Hamiltonian
\begin{equation}
    \hat{H} = \frac{1}{2}(\hat{p}^2 + \hat{x}^2).
    \label{eq:hamiltonian}
\end{equation}
Besides the standard position and energy-eigenstate bases for the Hilbert space (spanned by the states $|x\rangle$ and $|n\rangle$, respecitvely), one can also study the basis furnished by the \textit{coherent states}, or eigenbras of $\hat{a}^\dagger \equiv \frac{1}{\sqrt{2}}(-i\hat{p} + \hat{x})$ (see Appendix \ref{sec:Holomorphic_appendix} for details):
\begin{equation}
    \langle z| \hat{a}^\dagger = \langle z|z.
\end{equation}

Just as one can write down the position-space wavefunction for the state $|\psi\rangle$, $\psi(x) \equiv \langle x | \psi \rangle$, one can define coherent-state wavefunctions (frequently called \textit{holomorphic} wavefunctions)
\begin{equation}
    \psi(z) \equiv \langle z | \psi \rangle.
\end{equation}
In the quantum mechanics of the harmonic oscillator, these holomorphic wavefunctions are analogous to the wavefunctional $\Psi[A]$. For example, the holomorphic wavefunction of the position eigenstate $|x\rangle$ is
\begin{equation}
    \langle z | x \rangle = \pi^{-1/4} e^{\sqrt{2}zx - \frac{1}{2}x^2}.
\end{equation}
This last expression implies the following actions for the operators $\hat{x}$ and $\hat{a}$ on $\psi(z)$:
\begin{equation}
\begin{split}
    \sqrt{2}\hat{x} | \psi \rangle &\to \frac{\partial}{\partial z} \psi(z),
    \\
    \hat{a}^\dagger | \psi \rangle &\to z \psi(z).
    \label{eq:q_and_a_action}
\end{split}
\end{equation}

Importantly, since $\hat{a}^\dagger$ is not self-adjoint, the coherent states are not orthonormal. Despite this, it is not hard to show that they form a complete basis set (actually, as discussed in Appendix \ref{sec:Holomorphic_appendix}, the basis is overcomplete).

Of great relevance to this paper is the holomorphic inner product. As we will see below, the nonorthogonality of the coherent-state basis implies a nontrivial measure factor in this inner product. It is possible to derive the appropriate inner product using the known spectrum of the harmonic oscillator \cite{Bargmann1961,Bargmann1962}. However, since the Hamiltonian constraint (\ref{eq:hamiltonian_constaint}) does not take the form of a harmonic oscillator, we present here a more general derivation of the inner product. The crux of the proof is the \textit{reality conditions} obeyed by $\hat{x}$ and $\hat{a}^\dagger$:
\begin{equation}
    \hat{x}^\dagger = \hat{x}, \qquad \hat{a} + \hat{a}^\dagger = \sqrt{2} \hat{x}.
    \label{eq:SHO_reality_conditions}
\end{equation}
These conditions depend only on the operators $\hat{x}$ and $\hat{a}^\dagger$, \textit{not} on the structure of the Hamiltonian. Moreover, given the action of the operators $\hat{x}$ and $\hat{a}^\dagger$ on the holomorphic wavefunction $\psi(z)$, any purported inner product must satisfy these conditions. Exactly analogous conditions exist in canonical self-dual quantum gravity as discussed in Sec. \ref{sec:kodama}.

Making the ansatz
\begin{equation}
    \langle \psi | \phi \rangle = \int dz d\bar{z} \, e^{\mu(z, \bar{z})} \overline{\psi(z)} \phi(z)
    \label{eq:holomorphic_inner_product_2}
\end{equation}
and using the actions (\ref{eq:q_and_a_action}) of the $\hat{x}$ and $\hat{a}^\dagger$ operators, the first reality condition $\langle \psi | \hat{x} - \hat{x}^\dagger | \phi \rangle = 0$ becomes
\begin{equation}
\begin{split}
    \frac{1}{\sqrt{2}} \int dzd\bar{z} \, e^{\mu(z, \bar{z})} (\partial_z - \partial_{\bar{z}}) \overline{\psi(z)} \phi(z) = 0,
\end{split}
\end{equation}
where we used the holomorphicity of the functions $\psi(z)$. Integrating by parts we find, for all holomorphic functions $\psi$ and $\phi$,
\begin{equation}
\begin{split}
    \int dz d\bar{z} \, \left[(\partial_z - \partial_{\bar{z}}) e^{\mu(z, \bar{z})}\right] \overline{\psi(z)} \phi(z) = 0.
\end{split}
\end{equation}
A particular solution to this integral equation is found by letting the bracketed expression vanish; then $\mu(z, \bar{z})$ is a function of $z + \bar{z} = 2\Re z$ only.

The second reality condition yields
\begin{equation}
    \int dz d\bar{z} \, \left\{\left[z + \bar{z} + \partial_z\right] e^{\mu(z + \bar{z})}\right\} \overline{\psi(z)} \phi(z) = 0.
\end{equation}
Again, we let the bracketed expression vanish. To bolster the analogy with the self-dual variables, we solve this equation by Fourier expanding $e^{\mu}$:
\begin{equation}
    e^{\mu(\Re z)} = \int_{-\infty}^\infty dk \, e^{\Delta(k)} e^{-2i k \Re z}.
    \label{eq:measure_FT}
\end{equation}
The second reality condition is satisfied if we take $-i (2\Re z - i k) = \frac{\partial}{\partial k}[\Delta(k) - 2i k\Re z]$, for then
\begin{equation}
    \left[2 \Re z + \partial_z\right] e^{\mu(\Re z)} = i\int_{-\infty}^\infty dk \, \frac{\partial}{\partial k} e^{\Delta(k) - 2i k\Re z} = 0,
\end{equation}
where we have assumed $\lim_{k \to \pm \infty} e^{\Delta(k) - 2i k\Re z} = 0$ (this must be checked once $\Delta$ is known). Solving for $\Delta(k)$, substituting in (\ref{eq:measure_FT}), and integrating, we find
\begin{equation}
    e^{\mu(z, \bar{z})} = N_0 e^{-\frac{1}{2}(z + \bar{z})^2}
\end{equation}
where $N_0$ is some normalization. Taking $|\psi\rangle = |x\rangle$ and $| \phi \rangle = | y \rangle$ in (\ref{eq:holomorphic_inner_product_2}), we find $\langle x | y \rangle = \delta(x - y)$ if $N_0 = \frac{1}{4\pi i}$. It is easy to check directly that the resultant holomorphic inner product,
\begin{equation}
    \langle \psi | \phi \rangle = \int \frac{dz d\bar{z}}{4\pi i} \, e^{-\frac{1}{2}(z + \bar{z})^2} \overline{\psi(z)} \phi(z),
    \label{eq:holomorphic_inner_product}
\end{equation}
satisfies the reality conditions. 

At this point, two important notes are in order. First, although the value of $\langle \psi | \phi \rangle$ is uniquely determined (up to normalization) by the reality conditions, the overcompleteness of the coherent state basis means that the integral expression (\ref{eq:holomorphic_inner_product}) is nonunique; there exist many equivalent but distinct integral expressions for $\langle\psi|\phi\rangle$.

Second, and of great relevance to this paper, the holomorphic wavefunction $\psi(z)$ corresponding to a normalizable state $| \psi \rangle$ can be, and quite often is, nonnormalizable \emph{with respect to} the na\"{i}ve inner product. For example, with $|\psi\rangle = |n\rangle$, $\psi_n(z) = \frac{z^n}{\sqrt{n!}}e^{\frac{1}{2}z^2}$, which \textit{grows} exponentially with $|z|$ when $|\arg(z)| < \frac{\pi}{4}$. When $\psi_n(z)$ is substituted into the inner product (\ref{eq:holomorphic_inner_product}), however, the exponentials compensate each other and we obtain the correct result $\langle n | n \rangle = 1$. This shows the danger in drawing conclusions about the normalizability of connection-representation wavefunctionals $\Psi[A]$ (or any wavefunction written in terms of an overcomplete basis) when the correct inner product is not known. What we will soon see is that the inner product in the holomorphic representation of the CSK state exhibits a similar structure. Because this state of affairs recurs frequently throughout this article, we refer to states which are nonnormalizable with respect to the naive inner product as na\"{i}vely nonnormalizable. We emphasize that such states are not necessarily normalizable, but neither are they necessarily nonnormalizable.

\section{Chern-Simons-Kodama State}
\label{sec:kodama}

Having clarified the interpretation of connection-representation wavefunctionals, we are now in a position to discuss the Chern-Simons-Kodama state, which is an exact wavefunctional solution to the constraints of self-dual canonical quantum gravity in the connection representation.

In our discussion of the Hamiltonian formulation, we encountered the Hamiltonian constraint (\ref{eq:hamiltonian_constaint}). In the quantum theory, this constraint separates out the physical and unphysical states. There arises here an issue of operator ordering as the $\hat{E}^a_i(\vec{x})$ and $\hat{A}^i_a(\vec{x})$ operators are noncommuting. If we assume $EEF$ ordering, then in the connection representation the constraint becomes the \textit{Wheeler-DeWitt equation} (here $\ell_{\rm Pl}^2 \equiv 8\pi G \hbar$)
\begin{equation}
    \epsilon_{ijk} \frac{\delta}{\delta A_{ai}} \frac{\delta}{\delta A_{bj}} \left( F^k_{ab} + \ell_{\rm Pl}^2 \frac{\Lambda}{3} \epsilon_{abc} \frac{\delta}{\delta A_{ck}} \right)\Psi[A] = 0.
\end{equation}
In this form, it is manifest that a \textit{self-dual wavefunctional}, i.e. one satisfying
\begin{equation}
    \left( F^k_{ab} + \ell_{\rm Pl}^2\frac{\Lambda}{3} \epsilon_{abc} \frac{\delta}{\delta A_{ck}} \right)\Psi[A] = 0
\end{equation}
will automatically solve the Wheeler-DeWitt equation. The self-duality condition is straightforward to solve; its solution is the \textit{Chern-Simons-Kodama state}
\begin{equation}
    \Psi_{\rm K}[A] = N \exp\Bigg( \frac{3}{\ell_{\rm Pl}^2\Lambda} S_{\rm CS}[A] \Bigg).
    \label{eq:kodama_state}
\end{equation}
Here, we have introduced the \textit{Chern-Simons functional}
\begin{equation}
    S_{\rm CS}[A] = \int \text{Tr}\left( \frac{1}{2} A \wedge dA + \frac{1}{3} A \wedge A \wedge A \right).
\end{equation}
The existence of an exact solution to the Hamiltonian constraint is surprising and encouraging. However, the physical interpretation of this solution has been fraught with difficulties. Largely, these difficulties are centered on the normalizability of the Kodama state, which we now discuss.

\subsection{Normalizability and the Inner Product}\label{sec:norm_inner_product}

The Chern-Simons-Kodama state is na\"{i}vely nonnormalizable. In other quantum theories, where CSK-state analogs exist, this nonnormalizability excludes it from the physical Hilbert space. For example, the $\text{SU}(2)$ gauge theory Kodama-like state is \cite{Freidel:2003pu}
\begin{equation}
    \Psi_{\rm YM}[A] = \exp\left(\frac{1}{\hbar g^2} \int_\Sigma S_{\rm CS}[A]\right),
    \label{eq:Yang-Mills_Kodama}
\end{equation}
where $g$ is the Yang-Mills coupling. In Yang-Mills theory, the inner product is given by
\begin{equation}
    \langle \Psi_{\rm YM} \rvert \Psi_{\rm YM} \rangle = \int [dA] \, \lvert\Psi_{\rm YM}[A]\rvert^2.
    \label{eq:yang_mills_inner_product}
\end{equation}
If one decomposes $A$ in normal modes, one finds that $S_{\rm CS}$ is additive in independent modes, and that half of the modes enter into the sum with a positive sign, while half enter with a negative sign \cite{Witten:2003mb}. Therefore, whatever the sign of the exponent in (\ref{eq:Yang-Mills_Kodama}), the integral (\ref{eq:yang_mills_inner_product}) does not converge.

Na\"{i}vely, one would expect the same issue to arise for the gravitational Kodama state \footnote{Note, however, that it is not entirely clear in what sense one can define cosmological probabilities at all. Indeed, if there is only one universe, the frequentist interpretation of probability breaks down. In the absence of a probabilistic interpretation, it is no longer clear that normalizability should be taken as a foundational principle, and it may be that the nonnormalizable Kodama state is therefore not excluded from the physical Hilbert space.}. However, as we have seen, in quantum gravity the connection operator $\hat{A}^i_a$ is non-self-adjoint, so that the wavefunctional $\Psi[A]$ exists in a holomorphic representation and, by analogy with the holomorphic representation of the harmonic oscillator, the inner product is modified from (\ref{eq:yang_mills_inner_product}). In fact, classically the self-dual variables obey the reality conditions
\begin{equation}
    E^a_i = \bar{E}^a_i, \qquad A^i_a + \bar{A}^i_a = 2\Gamma^i_a(E)
    \label{eq:self-dual_reality_conditions}
\end{equation}
(here $\Gamma^i = -\frac{1}{2}\epsilon^{ijk}\omega_{jk}$ is the \textit{triad spin connection}), which must also be obeyed by the quantum operators when they are projected onto the physical Hilbert space. Note the similarity with the reality conditions (\ref{eq:SHO_reality_conditions}). If one assumes that the reality conditions are operator identities on the entire Hilbert space, then the inner product may be derived via procedure outlined in Sec. \ref{subsec:holomorphic_representation} (with minor modifications). This approach was taken in \cite{Alexander:2022ocp}, where the following inner product is proposed:
\begin{equation}
    \langle \Psi \rvert \Phi \rangle = \int [dA \, d\bar{A}] \overline{\Psi[A]} e^{-S(\Re A)} \Phi[A],
    \label{eq:4d_inner_product}
\end{equation}
with
\begin{equation}
    e^{-S(\Re A)} \equiv \int [dE] \exp\left( -\frac{1}{\ell_{\rm Pl}^2} \int e_i \wedge d_{\Re A}e^i \right),
    \label{eq:measure}
\end{equation}
where $d_{\Re A} = d + [A, \cdot]$ is a gauge covariant derivative and the commutator between two lie-algebra valued quantities $A$ and $B$ is $[A,B]^i = \epsilon^i_{\mathrm{\ }jk}A^j B^k$. We emphasize that, as in Sec. \ref{subsec:holomorphic_representation}, this inner product is \textit{not} a modification of the theory but is necessary to maintain consistency with the reality conditions (\ref{eq:self-dual_reality_conditions}). We also note, by analogy with the harmonic oscillator, that there are likely to be many equivalent forms of the inner product which, despite their equivalence, could look superficially quite different.

Prior to \cite{Alexander:2022ocp} and the present work it was unclear what the nontriviality of the gravitational inner product implied about the normalization of the Kodama state. Certainly the simple argument used for the Yang-Mills state does not apply when the inner product is nontrivial, but in the absence of an explicit expression for the inner product it was difficult to say much more.

The rest of this paper attempts to answer this outstanding question via a perturbative study of the inner product \eqref{eq:4d_inner_product} and its implications for the CSK state. In the remainder of this section, we briefly review earlier perturbative studies of the connection-representation inner product and the normalizability of the CSK state. In the next section, we use insight gained from these earlier studies to understand the linearization of (\ref{eq:4d_inner_product}) about de Sitter. We show in Sec. \ref{sec:linearization} how these de Sitter space expansions combine perturbatively with the inner product to render the CSK state normalizable for super-Planckian values of the cosmological constant.

\subsection{Gravitons and the Linearized Kodama State}

There have been some studies on the linearization of the CSK state in the literature \cite{Freidel:2003pu, Magueijo:2010ba}, investigating also the physical graviton states by means of a linearized inner product.
As in the nonperturbative regime (see discussion after Eq. \eqref{eq:self-dual_reality_conditions}), the linearized inner product is fixed by the reality conditions. Prior to imposing the reality conditions, the authors of \cite{Magueijo:2010ba} treated the densitized triad $E^a_i$ as complex, introducing independent graviton and antigraviton degrees of freedom into the theory. The resultant linearized inner product normalizes precisely half of the graviton modes, with the specific subset of normalized modes depending on operator ordering in the Hamiltonian constraint and whether the action (\ref{eq:sd_action_expanded}) is taken to be self-dual, anti-self-dual, or some linear combination of the two. For the self-dual theory and $EEF$ ordering (for which the Kodama state is an exact solution), Ref. \cite{Magueijo:2010ba} finds that it is left-handed gravitons and right-handed antigravitons that are normalizable.

More specifically, consider perturbations around de Sitter space in FLRW form. With the background metric
\begin{equation}
    \mathrm{d}s^2 = a(\eta)^2[-\mathrm{d}\eta^2 + (\delta_{ab} + h_{ab})\mathrm{d}x^a\mathrm{d}x^b], \quad h_{ab} \ll 1,
\end{equation}
the perturbations $\delta e^a_i$ of the densitized triad are defined by
\begin{equation}
    E^a_i = a^2\delta^a_i - a\delta e^a_i
    \label{eq:triad_perturb}
\end{equation}
while those of the self-dual connection satisfy
\begin{equation}
    A^i_a = iaH\delta^i_a + \frac{1}{a}\delta a^i_a.
    \label{eq:connection_perturb}
\end{equation}
In the first-order formalism of gravity, the fundamental variables are the tetrads \( e^I_{\ \mu} \), which relate to the metric via \( g_{\mu\nu} = e^I_{\ \mu} e^J_{\ \nu} \eta_{IJ} \). For FLRW backgrounds with scale factor \( a(\eta) \) in conformal coordinates, the spatial tetrad can be expanded as \( e^i_{\ a} = a\delta^i_{\ a} + \delta e^i_{\ a} \), where \( \delta e^i_{\ a} \) represents small perturbations encoding the tensor modes of gravity. These perturbations induce corresponding changes in the spatial metric up to second order via
\[
g_{ab} = e^i_{\ a} e^j_{\ b} \delta_{ij} = a^2 \delta_{ab} + a(\delta e_{ab} + \delta e_{ba}).
\]
Here, \( \delta e_{ab} \equiv \delta e^i_a \delta^j_b \delta_{ij} \) identifies the symmetric spatial components of the tetrad perturbation. In the transverse-traceless (TT) gauge, the physical metric perturbation is written $a^2 h_{ab}$. Comparing with the tetrad expansion, we find \( \delta e_{ab} = \frac{a}{2} h_{ab} \), so that the perturbation \( \delta e^a_{\ i} \) directly contains the graviton degrees of freedom in the first-order formalism.

The expansions of the perturbations of the dynamical variables are \cite{Magueijo:2010ba}
\begin{multline}
    \delta e_{ij}(\vec{x}) = \int\frac{d^3k}{(2\pi)^{3}}\sum_{r = \pm}\big[\epsilon^r_{ij}(\vec{k})\Psi_e(\vec{k},\eta)e_{r+}(\vec{k})e^{i\vec{k}\cdot\vec{x}} \\ + \epsilon^{r*}_{ij}(\vec{k})\Psi^*_e(\vec{k},\eta)e^\dagger_{r-}(\vec{k})e^{-i\vec{k}\cdot\vec{x}}\big]
    \label{eq:triad_expansion}
\end{multline}
for the triad and
\begin{multline}
    \delta a_{ij}(\vec{x}) = \int\frac{d^3k}{(2\pi)^{3}}\sum_{r = \pm}\big[\epsilon^r_{ij}(\vec{k})\Psi^{r+}_a(\vec{k},\eta)a_{r+}(\vec{k})e^{i\vec{k}\cdot\vec{x}} \\ + \epsilon^{r*}_{ij}(\vec{k})\Psi^{r-*}_a(\vec{k},\eta)a^\dagger_{r-}(\vec{k})e^{-i\vec{k}\cdot\vec{x}}\big]
    \label{eq:connection_expansion}
\end{multline}
for the connection. These are simply linearized spin-2 bosonic perturbations translated from the standard linearized metric and connection perturbations into the Ashtekar formalism. Here, the second superscript of $\Psi$ next to $r$ (denoted $p$ in what follows) labels graviton/antigraviton modes and $\epsilon^r_{ij}(\vec{k})$ are the two transverse, traceless, and symmetric polarization tensors. We work in the circular polarization basis, with
\begin{equation}
    \epsilon^{abc} k_b \epsilon^{r}_{cd}(\vec{k}) = irk\; \epsilon_{d}^{ra}(\vec{k}),
\end{equation}
$\epsilon^r_{ij}(\vec{k}) \epsilon^{s \;ij}(\vec{k}) = 0$, and $\epsilon^r_{ij}(\vec{k}) \epsilon^{s \;ij}(-\vec{k}) =  \epsilon^r_{ij}(\vec{k}) \left(\epsilon^{s \;ij}(\vec{k})\right)^* = 2 \delta^{rs}$.

The mode functions $\Psi_e$ corresponding to the Bunch-Davies vacuum are
\begin{equation}
    \Psi_e = \frac{e^{-ik\eta}}{\sqrt{2k}}\bigg(1-\frac{i}{k\eta}\bigg),
\end{equation}
and
\begin{equation}
    \Psi_a^{rp} = \left(ip \frac{\partial}{\partial \eta} + rk\right) \Psi_e.
    \label{eq:psis}
\end{equation}

Before we proceed, a word about the reality of the connection perturbations is in order. In the original formulation of the self-dual theory, the connection $A$ is allowed to be complex. However, solutions of the theory should satisfy reality conditions to make contact with classical (zero torsion) gravity. Indeed, a generic complex perturbation of the connection induces torsion, putting it at odds with classical general relativity. Restricting our perturbations to be along torsionless connections forces $\delta a^i_a$ to be real and to be relatable to the triad perturbations (this is the origin of equation (\ref{eq:psis}), and it explains the appearance of the exact same polarization tensors $\epsilon^r_{ij}$ in both mode decompositions (\ref{eq:triad_expansion}) and (\ref{eq:connection_expansion})). Since in this study we are interested in the normalizability of the graviton states, below we restrict ourselves to \textit{real}  $\delta a_{ia}(\vec{x})$.

In terms of the connection perturbations $\delta a$, the CSK state (\ref{eq:kodama_state}) is
\begin{equation}
    \Psi_{\rm K}[\delta a] = N \Psi_{\rm K}[A_{0}] \Psi_2[\delta a] \Psi_3[\delta a]
    \label{eq:kodama_state_perturbed}
\end{equation}
where $\Psi_i[\delta a] = \exp\left( \frac{3}{\ell_{\rm Pl}^2\Lambda} S^i_{\rm CS}[\delta a] \right)$ and $S^i_{\rm CS}[\delta a]$ is $\mathcal{O}(\delta a^i)$ (there is no $\Psi_1[\delta a]$ as de Sitter is an extremum of $\Psi_{\rm K}[A]$ under the na\"{i}ve inner product). Substituting the perturbations (\ref{eq:connection_perturb}) into the Chern-Simons functional and separating the terms by order, we find
\begin{equation}
\begin{split}
    S_{\rm CS}[A_{\rm dS}] &= 2ia^3H^3 \int d^3x = 2ia^3H^3 \delta^{3}(0),
    \\
    S^{(2)}_{\rm CS}[\delta a] &= -\frac{1}{a^2} \int d^3x \, \left( \epsilon^{abc} \delta a_{ia} \partial_b \delta a^i_c - iaH \delta a_{ia} \delta a^{ai} \right),
    \\
    S^{(3)}_{\rm CS}[\delta a] &= -\frac{1}{3a^3} \int d^3x \, \epsilon^{abc} \epsilon^{ijk} \delta a_{ia} \delta a_{jb} \delta a_{kc}.
\end{split}
\end{equation}
In terms of the modes (\ref{eq:connection_expansion}), the second-order term is
\begin{equation}
\begin{split}
    S^{(2)}_{\rm CS}[\delta a] &= \frac{2}{a^2} \int \frac{d^3k}{(2\pi)^3} \sum_r (iaH + rk) |\tilde{a}^r(\vec{k}, \eta)|^2
\end{split}
\end{equation}
where we have used the reality of $\delta a_{ia}(\vec{x})$ and defined
\begin{equation}
    \tilde{a}_{r\pm}(\vec{k}, \eta) \equiv \Psi^{r\pm}_a(\vec{k},\eta) a_{r\pm}(\vec{k})
\end{equation}
and
\begin{equation}
    \tilde{a}_{r}(\vec{k}, \eta) \equiv \tilde{a}_{r+}(\vec{k}, \eta) + \tilde{a}^*_{r-}(-\vec{k}, \eta).
\end{equation}
Thus, to \textit{linearized} order, the CSK state is
\begin{multline}
    \Psi_{\rm K}[\delta a] = N_0 \exp\Bigg[ \frac{6}{\Lambda l_{Pl}^2} \\ \times \int \frac{d^3k}{(2\pi)^3} \sum_r (iaH + rk) \Bigg|\frac{\tilde{a}^r(\vec{k})}{a(\eta)}\Bigg|^2 \Bigg].  \label{eq:kodama_state_perturbed_2}
\end{multline}
Since the real part of the coefficient of $a^r(\vec{k})$ depends on the label $r$, the CSK state is na\"{i}vely nonnormalizable at linear order (i.e. nonnormalizable with respect to the na\"ive inner product -- see the discussion at the end of Sec. \ref{sec:quantization}). In a theory where the inner product is (\ref{eq:yang_mills_inner_product}), such as $\text{SU}(2)$ gauge theory, this apparent nonnormalizability is the whole story, and we conclude that the CSK state is perturbatively nonnormalizable. However, since the inner product (\ref{eq:4d_inner_product}) for 4D quantum gravity is nontrivial, we cannot make definitive statements about the perturbative normalizability of the gravitational Chern-Simons-Kodama state before studying the linearization of the inner product measure. Therefore, in the following section, we use the perturbation expansions (\ref{eq:triad_expansion}) and (\ref{eq:connection_expansion}) to expand this measure about de Sitter space.

\section{Linearization of the Inner Product}
\label{sec:linearization}

In this section, we perform the linearization of the inner product (\ref{eq:4d_inner_product}). Since we are primarily interested in the normalizabiliy of the CSK state, our approach is to expand the connection $A^i_a$ about de Sitter space, where $\Psi_K[A]$ has a extremum with respect to the na\"{i}ve inner product. Then $\delta a$ in (\ref{eq:connection_perturb}) may be considered small, and since $\Re A_{dS} = 0$ we are justified in evaluating (\ref{eq:4d_inner_product}) as a perturbation series in $\Re A$. A careful analysis of the result, equation (\ref{eq:log_measure_final}), shows that it is convergent for super-Planckian cosmological constant. We end the section by discussing the regularization of the sub-Planckian divergence.

\subsection{Perturbation Series for $S(\Re A)$}

The inner product measure (\ref{eq:measure}) is defined in terms of a functional integral over the densitized triad $E^a_i$. To appropriately evaluate the measure, then, one should either express the integrand as a function of $E$ or change variables to the (nondensitized) frame field $e$ via (\ref{eq:densitized_triad}). However, here we make the approximation $[dE] \cong [de]$. Then the measure becomes Gaussian and can be integrated immediately (if formally):
\begin{equation}
\begin{split}
    e^{-S(\Re A)} &\cong \int [de] \exp\left( -\frac{1}{l_P^2} \int e_i \wedge d_{\Re A}e^i \right)
    \\
    &= \left[\det \left( \frac{2}{l_P^2} \wedge d_{\Re A} \right)\right]^{-1/2}
    \\
    &= \exp\left[ -\frac{1}{2} \text{Tr} \ln \left( \frac{2}{l_P^2} \wedge d_{\Re A} \right)\right]
    \label{eq:measure_explicit}
\end{split}
\end{equation}
Here, the determinant and trace are taken over \textit{all} indices, including spacetime ones. Before explicitly restricting to perturbations about de Sitter, it is useful to derive an exact, if formal, expression for the measure as a perturbation series in $\Re A$. To do this, we decompose $d_{\Re A} = d - [\cdot, \Re A]$, pull out a factor of $\frac{2}{l_P^2}\wedge d$, and Taylor expand the logarithm; the result is
\begin{equation}
    e^{-S(\Re A)} \propto \exp\left[ \frac{1}{2} \sum_{n = 1}^\infty \text{Tr} \left[\left( (\wedge d)^{-1} \wedge [\cdot, \Re A] \right)^n \right] \right]
\end{equation}
where the constant of proportionality is
\begin{equation}
    \mathcal{N}_0 = \left[\det \left( \frac{2}{l_P^2} \wedge d \right)\right]^{-1/2}.
\end{equation}
Thus, up to a constant,
\begin{equation}
\begin{split}
    S(\Re A) &= -\frac{1}{2} \sum_{n = 1}^\infty \frac{1}{n} \text{Tr} \left[\left( -(\wedge d)^{-1} \wedge [\Re A, \cdot] \right)^n \right].
    \label{eq:measure_exponent}
\end{split}
\end{equation}
The linearization of the inner product is found by computing the lowest-order terms in this series.

To evaluate this formal expression, let us first introduce the streamlined notation $\mathcal{D} \equiv \wedge d$, $\mathcal{L}_{\Re A} \equiv \wedge [\Re A, \cdot]$, and $\mathcal{D}_{\Re A} \equiv \mathcal{D} + \mathcal{L}_{\Re A}$. In terms of components,
\begin{equation}
\begin{split}
    \mathcal{D}^{ia \, jb}(\vec{x}, \vec{y}) &= l_P^{-2}\epsilon^{acb} \delta^{ij}\partial^{(y)}_c\ \delta^{(3)}(\vec{x} - \vec{y}),
    \\
    (\mathcal{L}_{\Re A})^{ia \, jb}(\vec{x}, \vec{y}) &= l_P^{-2}\epsilon^{acb} \epsilon^{ikj} \Re A_{kc}(\vec{x}) \delta^{(3)}(\vec{x} - \vec{y}).
    \label{eq:operators}
\end{split}
\end{equation}
Or, in momentum space,
\begin{equation}
\begin{split}
    \mathcal{D}^{ia \, jb}(\vec{k}, \vec{k}') &= -\frac{il_P^{-2}}{(2\pi)^3} \epsilon^{acb} \delta^{ij} k_c \delta^{(3)}(\vec{k} + \vec{k}'),
    \\
    (\mathcal{L}_{\Re A})^{ia \, jb}(\vec{k}, \vec{k}') &= \frac{l_P^{-2}}{(2\pi)^3}\epsilon^{acb} \epsilon^{ikj} \widetilde{\Re A}_{kc}(-\vec{k} - \vec{k}')
    \label{eq:D_and_L}
\end{split}
\end{equation}
(here a tilde denotes the Fourier transform). We see from these last expressions that $\mathcal{D} \sim \vec{k} \times$. Thus, $\mathcal{D}$ has a highly nontrivial kernel and is not a priori invertible (specifically, it annihilates all frame-fields longitudinal modes). However, with our perturbation expansions (\ref{eq:triad_expansion}) we chose a gauge in which the frame field perturbations are transverse, traceless, and symmetric (TTS). The functional integral (\ref{eq:measure_explicit}) is therefore restricted to the TTS subspace of field configuration space. For nonabelian gauge fields, such a subspace has nontrivial curvature, necessitating the introduction of Faddeev-Popov ghost fields. In our case, the gauge transformations of the metric are abelian, and the curvature introduced by restriction to TTS gauge is field-independent, contributing only an overall normalization.

Acting on transverse vectors, the cross product $\vec{k} \times$ squares to $-k^2$, so
\begin{equation}
\begin{split}
    (\mathcal{D}^2)^{ia \, jb}(\vec{k}, \vec{k}') &= \int d^3p \, \mathcal{D}^{ia \, kc}(\vec{k}, \vec{p}) \mathcal{D}_{kc}^{\mathrm{\ \ } jb}(\vec{p}, \vec{k}')
    \\
    &= -\frac{k^2}{(2\pi)^6l_P^4} \delta_{T}^{ab}(\vec{k}) \delta^{ij} \delta^{(3)}(\vec{k} - \vec{k}'),
    \label{eq:D_squared}
\end{split}
\end{equation}
where $\delta_T^{ab}(\vec{k})$ is the transverse delta function, i.e.
\begin{equation}
    \delta_T^{ab}(\vec{k}) \equiv \delta^{ab} - \frac{k^a k^b}{k^2}.
\end{equation}
Since (\ref{eq:D_squared}) is proportional to the identity, we see that $\mathcal{D}^{-1}$ is proportional to $\mathcal{D}$ in TTS gauge:
\begin{equation}
\begin{split}
    (\mathcal{D}^{-1})^{ia \, jb}(\vec{k}, \vec{k}') &= i\frac{(2\pi)^3 l_P^2}{k^2} \epsilon^{acb} \delta^{ij} k_c \delta^{(3)}(\vec{k} + \vec{k}').
    \label{eq:D_inverse}
\end{split}
\end{equation}

\subsection{Evaluation to Quadratic Order}
\label{subsection:measure_quadratic_order}

We are now in a position to evaluate the terms in the perturbation series (\ref{eq:measure_exponent}). Restoring indices, the trace $\text{Tr}[\mathcal{A}\mathcal{B}]$ of two operators $\mathcal{A}$ and $\mathcal{B}$ is
\begin{equation}
    \text{Tr}[\mathcal{A}\mathcal{B}] = \int \frac{d^3k}{(2\pi)^3} \, \frac{d^3p}{(2\pi)^3} \, \mathcal{A}^{ia \, jb}(\vec{k}, \vec{p}) \, \mathcal{B}^{jb \, ia}(\vec{p}, \vec{k}),
    \label{eq:trace}
\end{equation}
where summation is implied. Substituting the explicit expressions (\ref{eq:D_and_L}) and (\ref{eq:D_inverse}) into the $n = 1$ term in (\ref{eq:measure_exponent}), we find that the $\delta^{ij}$ appearing in $\mathcal{D}^{-1}$ is contracted with the $\epsilon^{ikj}$ appearing in $\mathcal{L}_{\Re A}$. Since the first is symmetric and the second is totally antisymmetric,
\begin{equation}
    \text{Tr}[\mathcal{D}^{-1}\mathcal{L}_{\Re A}] = 0.
\end{equation}

Moving on to the $n = 2$ term, we have, after some algebra,
\begin{equation}
\begin{split}
    \text{Tr}[(\mathcal{D}^{-1}\mathcal{L}_{\Re A})^2] = \int \frac{d^3k}{(2\pi)^3} \, \mathcal{T}^{ab}(\vec{k}) \widetilde{\Re A}_{ia}(\vec{k}) \widetilde{\Re A}^{i*}_b(\vec{k}),
    \label{eq:k_2_trace}
\end{split}
\end{equation}
where
\begin{equation}
\begin{split}
    \mathcal{T}^{ab}(\vec{k}) \equiv 2\int \frac{d^3p}{(2\pi)^3} \, \left[\frac{(p+k)^{(a} (p-k)^{b)}}{(\vec{p}+\vec{k})^2 (\vec{p}-\vec{k})^2}\right]
    \label{eq:T}
\end{split}
\end{equation}
and we have used the fact that $\widetilde{\Re A}^{ia}(-\vec{k}) = \widetilde{\Re A}^{ia*}(\vec{k})$, which holds for the Fourier modes of the real part of the connection. It can be seen by simple dimensional analysis that this integral is divergent and therefore must be regulated. We evaluate it explicitly, and discuss its regularization, in Appendix \ref{sec:appendix_T}. There, we show that 
\begin{equation}
\begin{split}
    \mathcal{T}^{ab}(\vec{k}) = \alpha |k| \left(\delta^{ab} + \frac{k^a k^b}{k^2}\right)
\end{split}
\end{equation}
where $\alpha$ is a divergent constant. After subtracting off the divergent part (see appendix \ref{sec:appendix_T} for details), we find
\begin{equation}
    \alpha = -\frac{1}{16}.
\end{equation}
In what follows, $\alpha$ is left as a variable to reflect the possibility of a different regularization scheme. Now, since we are working in TTS gauge, the $k^a k^b$ terms in $\mathcal{T}$ do not contribute to the integral (\ref{eq:k_2_trace}) except to pick out the contributions from the background connection. For de Sitter space, $\Re A_{dS} \equiv 0$ and these terms may be dropped entirely. Then (here $\lvert \widetilde{\Re A}(\vec{k}) \rvert^2 \equiv \widetilde{\Re A}_{ia}(\vec{k}) \widetilde{\Re A}^{ia*}(\vec{k})$)
\begin{equation}
\begin{split}
    \text{Tr}[(\mathcal{D}^{-1}\mathcal{L}_{\Re A})^2] = \alpha \int \frac{d^3k}{(2\pi)^3} \, k \lvert \widetilde{\Re A}(\vec{k}) \rvert^2
    \label{eq:answer}
\end{split}
\end{equation}
and the linearized measure is $e^{-S(\Re A)}$, where
\begin{equation}
\begin{split}
    S(\Re A) &= -\frac{\alpha}{4} \int \frac{d^3k}{(2\pi)^3} \, k \lvert \widetilde{\Re A}(\vec{k}) \rvert^2 + \mathcal{O}(A^4).
    \label{eq:log_measure_final}
\end{split}
\end{equation}
Or, in terms of the mode functions
\begin{equation}
    \begin{split}
        S(\Re A) = -\frac{\alpha}{4 a^2} \int \frac{d^3k}{(2\pi)^3} \, k & \sum_{r} \left|\tilde{a}_{r}(\vec{k}, \eta) + \tilde{a}^\dagger_{r}(-\vec{k}, \eta) \right|^2
        \\
        &\qquad\qquad\qquad+ \mathcal{O}(A^4).
        \label{eq:measure_modes}
    \end{split}
\end{equation}

\subsection{Linearized Norm}

In any quantum theory, the role of the inner product measure is to define the space of physical states by selecting the normalizable wavefunctionals. Crucially, it is not the measure or the wavefunctionals alone, but their interplay that determines whether the inner product  
\[
\langle \Psi | \Phi \rangle = \int [dA d\bar{A}]\, \overline{\Psi}[A]\, \Phi[A]
\]  
is well defined. In our context, we emphasize that the space of admissible states is always nonempty: there exist wavefunctionals for which the inner product converges.—e.g.,
\[
\Psi[A] = \Phi[A] = \chi[A]\, e^{S(\mathrm{Re}\,A)/2}
\]
for any functional $\chi[A]$ such that $\int [dAd\bar{A}] |\chi[A]|^2$ is well defined. This illustrates that the measure, suitably paired with appropriate wavefunctionals, can render the Hilbert space structure nontrivial and physically meaningful, at least after an appropriate regularization procedure.

The central question, then, is not whether normalizable wavefunctionals exist, but whether any exist which also satisfy the physical constraints of the theory. In particular, we ask whether the linearized CSK state --- which is already known to obey the constraints --- constitutes such a wavefunctional. We now turn to this question.

Combining the linearization of the measure {\eqref{eq:log_measure_final}} with the linearization of the Kodama state \eqref{eq:kodama_state_perturbed_2}, we compute the perturbative norm of the Kodama state:
\begin{multline}
    \langle\Psi|\Psi\rangle = \mathcal{N}\int [da d\bar{a}] \\ \times \exp\Bigg\{-\frac{1}{2}\int \frac{d^3k}{(2\pi)^3} \sum_r k\left(-\alpha - \frac{24r}{\ell_{\rm Pl}^2 \Lambda} \right)\left|\frac{\tilde{a}_r(\vec{k},\eta)}{a(\eta)}\right|^2  \Bigg\}
    \\
    = \mathcal{N} \langle\Psi|\Psi\rangle_+ \langle\Psi|\Psi\rangle_-\hspace{33.1mm}
    \label{eq:kodama_norm}
\end{multline}
where
\begin{multline}
    \langle\Psi|\Psi\rangle_r = \int [da_r d\bar{a}_r] \\ \times \exp\Bigg\{-\frac{1}{2}\int \frac{d^3k}{(2\pi)^3} k\left(-\alpha - \frac{24r}{\ell_{\rm Pl}^2 \Lambda} \right)\left|\frac{\tilde{a}_r(\vec{k},\eta)}{a(\eta)}\right|^2  \Bigg\}.
    \\
\end{multline}
One interesting property of this norm is that it manifestly \textit{factorizes} into two pieces, one for each graviton chirality. For $\alpha = -\frac{1}{16}$, the right-handed ($r = -$) piece is always convergent, while the left-handed ($r = +$) piece is normalizable so long as
\begin{equation}
    -\alpha\Lambda > \frac{24}{l_{Pl}^2}.
\label{eq:chiral-normalizability}
\end{equation}
For $\alpha = -\frac{1}{16}$, this is satisfied when
\begin{equation}
    \Lambda > \frac{384}{l_{Pl}^2} \equiv \Lambda_c.
\end{equation}
Therefore, we find that there is a super-Planckian critical value $\Lambda_c$ of the cosmological constant, above which the Kodama state is perturbative normalizable but below which precisely \textit{half} of the modes (those with $r = +$) are normalizable. (In a sense, then, one \textit{chiral sector} of the Kodama state is normalizable).

Despite its nonnormalizability, it is still possible to find meaningful interpretations of the $\Lambda \leq \Lambda_c$ Kodama state as a normalizable wavefunction. We may, for example, attempt normalization via superselection. Namely, the normalizable modes may be superselected and the nonnormalizable modes discarded. This is similar to the route suggested in \cite{Magueijo:2010ba}, though the specific modes which must be discarded here cannot be neatly distinguished according to graviton/antigraviton and right-handed/left-handed as they were there due to the presence of the $\alpha$ term.

In the following section, we will discuss an idea to make the $\Lambda \leq \Lambda_c$ Chern-Simons-Kodama state perturbatively normalizable. Namely, we will discuss a Wick-like transformation that, when applied to the CSK state, ensures that all of the graviton modes contribute to normalizability.

\subsection{The Norm of the Perturbative Kodama State}

As noted above, the integral (\ref{eq:kodama_norm}) is only well-defined when $\Lambda > \Lambda_c$. What, though, of the normalizability \textit{below} $\Lambda_c$? Then half of the modes are \textit{nonnormalizable} in the sense that they contribute an indefinite factor to the norm \eqref{eq:kodama_norm}. We can understand what is happening here with a simple example: consider the one-dimensional Gaussian integral
\begin{equation}
    \int_{-\infty}^\infty dx \, e^{k x^2}
    \label{eq:gaussian_example}
\end{equation}
(at second order in $\Re{A}$, the divergence is precisely a multidimensional generalization of this example). This integral is divergent, but if it is evaluated by analytic continuation we find the finite answer $\sqrt{\frac{\pi}{-k}}$. This procedure is commonly used in the evaluation of nonrelativistic single-particle propagators. In effect, it involves the replacement $x \to ix$, which changes the sign in the exponent at the cost of introducing a factor of $i$. The linearized norm studied here consists of functional integrals (some of which are divergent) over a Gaussian functional $\exp[-\frac{1}{2}\vec{a}^\dagger \mathcal{Q} \vec{a}]$ for some operator $\mathcal{Q}$ (see Eq. \eqref{eq:kodama_norm}). Thus, if there was a self-consistent and physically meaningful way to make a replacement analogous to $x \to ix$ in the negative-eigenvalue modes of $\mathcal{Q}$, convergence could be assured for any regularized $\alpha$.

Motivated by this example, we propose that the nonnormalizability of the Kodama state can be resolved by the use of Thiemann's phase-space Wick rotation \cite{Thiemann:1995ug, Ashtekar:1995qw}. Thiemann proves that the symplectic structure on the space of Ashtekar variables is preserved by a canonical transformation on the phase space. This is accomplished by a canonical transformation generated by a \textit{complexifier function} \( C \), typically defined as
\[
C := \int_{\Sigma} d^3x \, K^i_a E^a_i,
\]
where \( K^i_a \) is the extrinsic curvature expressed in the triad basis and \( \Sigma \) is a spatial hypersurface. This can be understood as a flow on the space of connections generated by the complexifier generator $C$.

The complexifier generates a one-parameter family of complex canonical transformations acting on phase space functions via the exponential of the Poisson bracket:
\[
A^i_a \mapsto W_\theta A^i_a := e^{i \theta \{ \cdot, C \}} A^i_a
\]
where the exponential of the Poisson bracket denotes repeated evaluation of the Poisson bracket acting on the connection in accordance with the Taylor expansion of the exponential function. This transformation rotates the connection in the complexified phase space. For \( \theta = \pm \frac{\pi}{2} \), the rotation maps the real-valued Ashtekar-Barbero \cite{Ashtekar:1986yd, Ashtekar:1987gu, BarberoG:1994eia} connection to the (anti-)self-dual connection, which is complex \cite{Thiemann:1995ug, Ashtekar:1995qw}:
\begin{equation*}
W_{\pm \frac{\pi}{2}}A^i_a = \Gamma^i_a \pm i K^i_a.
\end{equation*}
Thus, the phase-space Wick rotation is a complex canonical transformation $W$ that interpolates between the real and complex formulations of the self-dual connection, while preserving the canonical Poisson structure. By construction, $W$ preserves the Poisson bracket structure of the canonical theory and, of course, an analogous operation exists on the quantum theory. In both cases, the name \textit{Wick rotation} is applicable since $W$ maps between Euclidean and Lorentzian theories.

To perform this transformation on the perturbed connection, we begin by defining the operator
\begin{equation}\label{eq:NewThiemannW}
    \hat{W} = \exp\Big[-\frac{\pi}{2} \hat{C} \Big],
\end{equation}
where $\hat{C}$ is the infinitesimal generator
\begin{equation}\label{eq:NewThiemannC}
    \hat{C} = \int \frac{d^3k}{(2\pi)^3} \hat{\tilde{a}}_+ (\vec{k}, \eta) \hat{\tilde{e}}^+(\vec{k}, \eta).
\end{equation}
Here $\hat{\tilde{a}}_+ (\vec{k}, \eta)$ is the operator corresponding to the connection mode $\tilde{a}_+(\vec{k}, \eta)$ and $\hat{\tilde{e}}^+(\vec{k}, \eta)$ is the operator corresponding to the triad mode $\tilde{e}^+(\vec{k}, \eta)$ which is canonically conjugate to $\tilde{a}_+(\vec{k}, \eta)$. When this operator acts on the CSK state $| \Psi_K \rangle$, its effect is to Wick rotate precisely those modes $\tilde{a}_r(\vec{k}, \eta)$ for which $\left(-\alpha - \frac{24r}{\ell_{\rm Pl}^2 \Lambda} \right) 
< 0$. All other modes are left invariant. That is, $\hat{W}$ induces a map $\mathcal{W} : \tilde{a}_r(\vec{k}, \eta) \to i^{\delta_{r}^1} \tilde{a}_r(\vec{k}, \eta)$ on the perturbation modes.

The  Wick rotation $\hat{W}$ is similar to Thiemann's original form in that we are still transforming variables in the phase space, but it is different in that it acts trivially on half of the modes of the Ashtekar connection. As a result, $\hat{W}$ leaves the Ashtekar connection complex. However, in general (and certainly in this case) $\hat{W}$ does \textit{not} commute with the Hamiltonian constraint $\hat{\mathcal{H}}$, so that the state $\hat{W}| \Psi_K \rangle$ becomes a solution to a modified constraint. In spite of this fact, the benefit of introducing $\hat{W}$ is the following map on the graviton modes:
\begin{equation}
    \mathcal{W} : |\tilde{a}_r(\vec{k}, \eta)|^2 \to -r|\tilde{a}_r(\vec{k}, \eta)|^2
\end{equation}
so that (here $| \Psi_K \rangle_W \equiv \hat{W} | \Psi_K \rangle$)
\begin{multline}
    _W\langle\Psi|\Psi\rangle_W = \mathcal{N}\int [da d\bar{a}] \\ \times \exp\Bigg\{-\frac{1}{2}\int \frac{d^3k}{(2\pi)^3} \sum_r k\left(\alpha r + \frac{24}{\ell_{\rm Pl}^2 \Lambda} \right)\left|\frac{\tilde{a}_r(\vec{k},\eta)}{a(\eta)}\right|^2  \Bigg\}.
    \label{eq:ThiemannKodamaNorm}
\end{multline}
which is convergent for $\Lambda < \Lambda_{c}$.

Here we summarize the previous steps running from equation \eqref{eq:kodama_norm} to \eqref{eq:ThiemannKodamaNorm}. First, one finds the norm of the Chern-Simons-Kodama state to quadratic order. This is expressed as a Gaussian integral explicitly in equation \eqref{eq:kodama_norm}. A phase-space Wick rotation given by \eqref{eq:NewThiemannW} and \eqref{eq:NewThiemannC} is applied to the CSK state. Finally, the explicit evaluation of the norm after the Wick rotation is in equation \eqref{eq:ThiemannKodamaNorm}.

\section{Discussion}
\label{sec:discussion}

We have shown that the perturbative Kodama state is normalizable for super-Planckian $\Lambda$ and have suggested a method (Thiemann's phase-space Wick rotation) for rendering it fully normalizable in the regime of small cosmological constant. However, it may be possible to address the small-$\Lambda$ divergences without the need for Wick rotation. Indeed, these divergences arise in the linearized expansion under two assumptions: (1) that de Sitter space is flatly sliced, and (2) that the functional measures $[de]$ and $[dE]$ can be treated as equivalent to second order in $\Re A$. Each of these assumptions, if relaxed, opens potential avenues for regulating the perturbative norm.

The divergence associated with $\alpha$ appears to stem from two intertwined issues in the flat slicing: the infinite volume of $\mathbb{R}^3$ and the fact that the gauge-fixed propagator relies on the differential operator $\wedge d$, which is not elliptic on $\mathbb{R}^3$. A natural resolution is to reperform the perturbative expansion on \emph{global} de Sitter space, where the spatial slices are compact $S^3$ manifolds. On $S^3$, the operator $\wedge d$ becomes elliptic, and its inverse $\mathcal{D}^{-1}$ can be rigorously defined through spectral methods. The mode expansions then involve discrete eigenvalues and vector spherical harmonics, ensuring that the trace 
\[
\mathrm{Tr} \log(1 - \mathcal{D}^{-1}[\Re A, \cdot])
\]
is well defined and free from infrared divergences. In this formulation, $\alpha$ is naturally regularized, and the entire perturbative expansion becomes resummable.

In parallel, the approximation $[dE] \cong [de]$ warrants further scrutiny. Since the densitized triad is defined as $E = (\det e) e^{-1}$, its variation satisfies
\[
dE^a_i = \epsilon^{abc} \epsilon_{ijk} e_b^j \, de_c^k \neq de^a_i,
\]
so that the Jacobian between $[dE]$ and $[de]$ is nontrivial. While our results hold at quadratic order under this approximation, the full functional Jacobian could influence higher-order corrections and potentially offer an alternative regularization mechanism for the divergent term $\alpha$. It remains an open question whether a more faithful treatment of the measure can render the norm finite without resorting to compactification.

Importantly, these two approaches—compactifying the spatial slice and refining the path integral measure—are not mutually exclusive. Their combination may be necessary to achieve full control over the perturbative expansion.

More broadly, it is worth reflecting on whether perturbative nonnormalizability at quadratic order should be regarded as a fundamental pathology. In fact, there are instructive analogies from quantum mechanics. For instance, consider the wavefunction
\[
\psi(x) = \exp\left( \sqrt{\frac{\beta}{\hbar}} x^2 - \frac{\beta}{3\hbar} x^4 \right),
\]
which is an exact energy eigenstate of the Hamiltonian
\[
H = -\frac{\hbar^2}{2m} \frac{d^2}{dx^2} + \frac{\beta^2}{2m} x^6 - \frac{\hbar \beta^{3/2}}{2m} x^4.
\]
Here, truncating the wavefunction to its quadratic behavior yields a divergent norm, but the full wavefunction is perfectly normalizable. This suggests that nonnormalizability at leading order can be an artifact of truncation and that full nonperturbative structures may resolve such divergences.

In our context, we have already shown that a specified Wick rotation can render the full perturbative state normalizable. This provides a rare example in quantum gravity where the normalizability of a vacuum wavefunctional is explicitly recovered via a controlled analytic continuation. 

We conjecture that this approach may generalize beyond the CSK state and offer a template for constructing viable vacuum states in nonperturbative quantum gravity with $\Lambda > 0$. The extension of our holomorphic inner product to compact spatial topologies and higher-order terms will be the focus of future work.

\section{Conclusion}
\label{sec:conclusion}

One of the main goals of this paper was to clarify the coherent state formalism in the holomorphic representation. In particular, we noted that the na\"{i}ve inner product is inconsistent with the self-dual formalism and showed that the nonperturbative inner product proposed in \cite{Alexander:2022ocp} is equivalent to the unique inner product for holomorphic wavefunctionals.

A second goal was a perturbative analysis of both the Chern-Simons-Kodama state and its inner product to quadratic order around the flat slicing of de Sitter space. After subtracting off the divergent part of a constant $\alpha$ which emerges from the perturbative analysis of the inner product measure, we find that the inner product at quadratic order is nonnormalizable for values of the cosmological constant $\Lambda$ which lie below some super-Planckian critical value $\Lambda_c$. This is because (for $\Lambda < \Lambda_c$) half of the perturbation modes appear in the norm $\langle \Psi_K | \Psi_K \rangle$ with a positive coefficient. However, for $\Lambda > \Lambda_c$ the sign of these modes reverses and the Chern-Simons-Kodama state is pertubatively normalizable. 

We posit that the divergent part of $\alpha$ is an artifact of the unbounded spatial domain of the flat slicing of de Sitter space and an improper treatment of the $\wedge d$ operator. When the theory is properly formulated on compact spatial slices, we expect that the series expansion of the inner product will resum in a manner analogous to known examples in QFT.

We further suggest that the introduction of Thiemann's transformation, which (despite not commuting with the Hamiltonian constraint) effectively Wick rotates the problematic perturbation modes while leaving the others invariant, could remedy the nonnormalizability for $\Lambda < \Lambda_c$ by changing the problematic signs in our expression for $\langle \Psi_K | \Psi_K \rangle$.

These findings open the door to a consistent semiclassical limit of the CSK state and motivate future work on global de Sitter quantization and higher-order corrections to the inner product.

\acknowledgments
We would like to thank João Magueijo, Laurent Freidel, Wolfgang Wieland, Lee Smolin, David Spergel, Rob Myers, Jim Gates, Michael Peskin, Bruno Alexandre, Etera Livine, Daine Danielson, Ibou Bah, Cyril Creque Sarbinowski and Antonino Marciano for discussions.

\begin{appendices}
\section{Holomorphic Representation of the Harmonic Oscillator}
\label{sec:Holomorphic_appendix}

In this section, we look more closely at the holomorphic representation, also called the Segal-Bargmann representation, of the harmonic oscillator \cite{Bargmann1961,Bargmann1962, Itzykson:1980rh}, providing some details which were glossed over in the main text. As there, we begin with the quantum harmonic oscillator with $m = \omega = 1$ and Hamiltonian
\begin{equation}
    \hat{H} = \frac{1}{2}(\hat{p}^2 + \hat{x}^2).
    \label{eq:app_hamiltonian}
\end{equation}
A general quantum state $|\psi\rangle$ of this system can be written as some function $f_\psi$ of the creation operator $\hat{a}^\dagger \equiv \frac{1}{\sqrt{2}}(-i\hat{p} + \hat{x})$ acting on the vacuum state $|0\rangle$:
\begin{equation}
    |\psi\rangle = f_{\psi}(\hat{a}^\dagger)|0\rangle.
\end{equation}
There is a close relationship between the function $f_\psi(\hat{a}^\dagger)$ and the coherent states. Indeed, the corresponding kets $|z\rangle$ are constructed from the creation operator as
\begin{equation}
    |z\rangle = \bar{N}_z e^{\bar{z}\hat{a}^\dagger}|0\rangle,
\end{equation}
where $N_z \equiv N_z(z, \bar{z})$ is some normalization. There are many useful choices for $N_z$, all related by a simple rescaling of the states $| z \rangle$. Here we follow Ref. \cite{Kodama:1990sc} in taking $N_z = e^{\frac{1}{2}z^2}$; then Eqs. (\ref{eq:coherent_state_FT})---(\ref{eq:coherent_state_expansion}) become particularly simple.

The overlap between a coherent state and a general state $|\psi\rangle$ is
\begin{equation}
    \langle z|\psi\rangle = \langle z|f_{\psi}(\hat{a}^\dagger)|0\rangle = f_{\psi}(z) \langle z|0\rangle = f_{\psi}(z)e^{\frac{1}{2}z^2}.
    \label{eq:app_holomorphic-wavefunction}
\end{equation}
The holomorphic function $F_\psi(z) \equiv e^{\frac{1}{2}z^2}f_{\psi}(z)$ is thus a wavefunction in a \textit{coherent state basis}, analogous to the wavefunctional $\Psi[A]$. In particular, the holomorphic wavefunction of the position eigenstate $|x\rangle$ is
\begin{equation}
    \langle z | x \rangle = \pi^{-1/4} e^{\sqrt{2}zx - \frac{1}{2}x^2}.
\end{equation}
By inserting a resolution of the identity in $\langle z | \psi \rangle$, we can write $F_{\psi}(z)$ in terms of the position-representation wavefunction $\psi(x)$:
\begin{equation}
    F_\psi(z) = \pi^{-1/4} \int_{-\infty}^\infty dx \, e^{-\frac{1}{2}x^2 + \sqrt{2}zx} \psi(x).
    \label{eq:coherent_state_FT}
\end{equation}

Since $\hat{x}$ acts on $\psi(x)$ via multiplication by $x$ and since $\langle z|$ is an eigenstate of $\hat{a}^\dagger$, this last expression implies
\begin{equation}
\begin{split}
    \sqrt{2}\hat{x} | \psi \rangle &\to \frac{\partial}{\partial z} F_\psi(z),
    \\
    \hat{a}^\dagger | \psi \rangle &\to z F_\psi(z).
    \label{eq:app_q_and_a_action}
\end{split}
\end{equation}

Since $\hat{a}^\dagger$ is not self-adjoint, the coherent states are not orthonormal:
\begin{equation}
    \langle z | z' \rangle = N_z \bar{N}_{z'} e^{z\bar{z}'}.
\end{equation}
Despite this, it follows from the transformation (\ref{eq:coherent_state_FT}) that they form a complete basis set (actually, the basis is overcomplete). Multiplying by $e^{-\sqrt{2}zy}$, integrating along the imaginary axis of $z$, and solving for $\psi(y)$,
\begin{equation}
    \psi(y) = (4\pi)^{1/4} e^{\frac{1}{2}y^2} \int_{-\infty}^\infty d\Im z \, e^{-\sqrt{2}zy} F_\psi(z).
    \label{eq:coherent_state_IFT}
\end{equation}
Since $\psi(x)$ can always be reconstructed from $F_\psi(z)$ via this expression, the $| z \rangle$ basis spans the Hilbert space. Indeed, from (\ref{eq:coherent_state_IFT}) we derive the explicit expansion
\begin{equation}
    \langle x | = (4\pi)^{1/4} e^{\frac{1}{2}x^2} \int d\Im z \, e^{-\sqrt{2}zx} \langle z |.
    \label{eq:coherent_state_expansion}
\end{equation}
which can be used to transform between the coherent-state basis and the position-state basis at will.

One often sees the holomorphic inner product presented in terms of the functions $f_\psi$ in the form
\begin{equation}
    \langle\phi|\psi\rangle = \int \frac{dzd\bar{z}}{2\pi i} e^{-|z|^2}\overline{f_\phi(z)} f_\psi(\bar{z}).
    \label{eq:holomorphic_inner_product_1}
\end{equation}
By equation (\ref{eq:app_holomorphic-wavefunction}), this is equivalent to the form \eqref{eq:holomorphic_inner_product} presented in Sec. \ref{subsec:holomorphic_representation}.

\section{Evaluation of $\mathcal{T}^{ab}$}
\label{sec:appendix_T}

In Sec. \ref{subsection:measure_quadratic_order} we cited the result
\begin{equation}
    \mathcal{T}^{ab}(\vec{k}) = -\frac{1}{16}k \Big(\delta^{ab} + \frac{k^a k^b}{k^2}\Big),
\end{equation}
where $\mathcal{T}^{ab}(\vec{k})$ is defined by the integral (\ref{eq:T}), without proof. The purpose of this appendix is to show how this was derived.

We wish to compute the following integral:
\begin{equation}
    \mathcal{T}^{ab}(\vec{k}) = \frac{2}{(2\pi)^3}\int d^Dp \frac{(p+k)^{(a}(p-k)^{b)}}{(p+k)^2(p-k)^2}.
\end{equation}
First we will make the integration variable substitution $l = p-k$ and define
\begin{equation}
\begin{split}
    I^{ab}(\vec{q}) &= \int d^Dl \frac{l^a l^b}{(l+q)^2 l^2}
    \\
    I^b(\vec{q}) &= \int d^Dl \frac{l^b}{(l+q)^2 l^2}.
\end{split}
\end{equation}
Then we can write the original integral as the sum
\begin{equation}
    \mathcal{T}^{ab}(\vec{k}) = \frac{2}{(2\pi)^3} I^{(ab)}(2\vec{k}) + \frac{2}{(2\pi)^3} (2k)^{(a}I^{b)}(2\vec{k}).
    \label{eq:T_from_Is}
\end{equation}

To evaluate $I^b$, we note, by rotational invariance, that
\begin{equation}
    I^b = q^b I
\end{equation}
where
\begin{equation}
    I = \frac{1}{q^2}q_b I^b = \frac{1}{q^2}\int d^Dl \frac{q_b l^b}{(l+q)^2 l^2}.
    \label{eq:ql_identity}
\end{equation}
Then, using the identity
\begin{equation}
    (l+q)^2 - l^2 - q^2 = 2q_b l^b
\end{equation}
we can write
\begin{equation}
    I = \frac{1}{2q^2}\int d^Dl \Big[\frac{1}{l^2} - \frac{1}{(l+q)^2} - \frac{q^2}{(l+q)^2 l^2}\Big].
\end{equation}
The second term cancels the first, as can be seen with the integration variable substitution $l' = l+q$. Thus,
\begin{equation}
    I^b(\vec{q}) = -\frac{q^b}{2}\int d^Dl \frac{1}{(l+q)^2 l^2}.
\end{equation}

Now let us turn to $I^{ab}(\vec{q})$. By rotational invariance and the fact that the integrand in $I^{ab}(\vec{q})$ is an even function of $l$, we can write
\begin{equation}
    I^{ab}(\vec{q}) = q^a q^b I_1(\vec{q}) + \delta^{ab}I_2(\vec{q})
    \label{eq:Iab}
\end{equation}
where
\begin{equation}
    I_1(\vec{q}) = -\frac{1}{(D-1)q^2}\Big(\delta_{ab} - \frac{Dq_a q_b}{q^2}\Big)I^{ab}(\vec{q})
\end{equation}
and
\begin{equation}
    I_2(\vec{q}) = \frac{1}{D-1}\Big(\delta_{ab} - \frac{q_a q_b}{q^2}\Big)I^{ab}(\vec{q}).
\end{equation}
Therefore, we must compute the trace
\begin{equation}
    \delta_{ab}I^{ab}(\vec{q}) = \int d^Dl \frac{l^2}{(l+q)^2 l^2} = \int d^Dl \frac{1}{l^2}.
\end{equation}
and the product $q_a q_b I^{ab}(\vec{q})$. Writing
\begin{equation}
    q_a I^{ab}(\vec{q}) = \int d^Dl \frac{q_a l^a l^b}{(l+q)^2 l^2}
\end{equation}
and using again the identity (\ref{eq:ql_identity}) and the change of variables $l' = l + q$, we obtain
\begin{multline}
    q_a I^{ab}(\vec{q}) = \frac{q^b}{2}\int d^Dl \frac{1}{l^2} - \frac{q^2}{2}\int d^Dl \frac{l^b}{(l+q)^2 l^2} \\ = \frac{q^b}{2}\int d^Dl \frac{1}{l^2} - \frac{q^2}{2}I^b(\vec{q}).
\end{multline}
Therefore,
\begin{equation}
    q_a q_b I^{ab}(\vec{q}) = \frac{q^2}{2}\int d^Dl \frac{1}{l^2} - \frac{q^2}{2}q_b I^b(\vec{q}).
\end{equation}
Substituting these results into (\ref{eq:Iab}), we find
\begin{multline}
    I^{ab}(\vec{q}) = \Big[\frac{D-2}{2(D-1)q^2}q^a q^b + \frac{1}{2(D-1)}\delta^{ab}\Big]\int d^Dl \frac{1}{l^2} \\ + \Big[-\frac{D}{2(D-1)q^2}q^a q^b + \frac{1}{2(D-1)}\delta^{ab}\Big]q_c I^c(\vec{q})
\end{multline}

It remains to evaluate $I^b(\vec{q})$ explicitly. Using Feynman integration tricks and completing the square in the denominator, we come to
\begin{equation}
    \int d^Dl \frac{1}{(l+q)^2 l^2} = \int d^Du \int_0^1 dx \frac{1}{[u^2 + x(1-x)q^2]^2}.
\end{equation}
Via the formula (see e.g. \cite{Nastase:2019,Schwartz:2014sze}):
\begin{equation}
    \int \frac{d^Dl}{(2\pi)^D} \frac{1}{(l^2 + x)^2} = \frac{1}{(4\pi)^{D/2}}\Gamma\Big(2-\frac{D}{2}\Big)x^{\frac{D}{2}-2}
\end{equation}
we find
\begin{equation}
\begin{split}
    &\int d^Dl \frac{1}{(l+q)^2 l^2} = \int_0^1 dx \frac{(2\pi)^D}{(4\pi)^{D/2}}\Gamma\Big(2-\frac{D}{2}\Big)\big(x(1-x)q^2\big)^{\frac{D}{2}-2}
    \\
    &= \frac{(2\pi)^D}{(4\pi)^{D/2}}\Gamma\Big(2-\frac{D}{2}\Big)(q^2)^{\frac{D}{2}-2}\frac{2^{3-D}}{\Gamma\Big(\frac{D-1}{2}\Big)}\sqrt{\pi}\Gamma\Big(\frac{D}{2}-1\Big).
\end{split}
\end{equation}
Or, in the limit that the number of spatial dimensions $D \to 3$
\begin{equation}
    \lim_{D\rightarrow 3}\int d^Dl \frac{1}{(l+q)^2 l^2} = \frac{\pi^3}{q}.
\end{equation}
Therefore,
\begin{equation}
    I^b(\vec{q}) = -\frac{\pi^3 q^b}{2q} \implies q_b I^b(\vec{q}) = -\frac{\pi^3}{2}q.
\end{equation}
The remaining integral appearing in $I^{ab}(\vec{q})$ is divergent. However, we can quantify its divergence by taking a limit:
\begin{equation}
    \int d^Dl \frac{1}{l^2} = \frac{2\pi^{D/2}}{\Gamma\Big(\frac{D}{2}\Big)}\int_0^\infty dl \ l^{D-3}
\end{equation}
so 
\begin{equation}
    \lim_{D\rightarrow 3}\int d^Dl \frac{1}{l^2} = 4\pi q\lim_{D\rightarrow 3}\int_0^\infty d\tilde{l} \ \tilde{l}^{D-3},
\end{equation}
where we rescaled the integration variable to $\tilde{l} = l/q$ in order to get a dimensionless integral. Substituting these results into $I^{ab}$, we find
\begin{multline}
    I^{ab}(\vec{q}) = \frac{1}{4}\Big(\delta^{ab} + \frac{q^a q^b}{q^2}\Big)\Big(4\pi q\lim_{D\rightarrow 3}\int_0^\infty dl \ l^{D-3}\Big) \\ + \frac{1}{4}\Big(\delta^{ab} - \frac{3q_a q_b}{q^2}\Big)\Big(-\frac{\pi^3}{2}q\Big)
\end{multline}
Therefore, from (\ref{eq:T_from_Is}),
\begin{equation}
    \mathcal{T}^{ab}(\vec{k}) = \frac{1}{2(2\pi)^3}\Big[8\pi k \lim_{D\rightarrow 3}\int_0^\infty dl \ l^{D-3} - k\pi^3 \Big]\Big(\delta^{ab} + \frac{k^a k^b}{k^2}\Big).
\end{equation}

Note that the constant of proportionality between $\mathcal{T}^{ab}$ and $\left(\delta^{ab} + \frac{k^ak^b}{k^2}\right)$ is $(\Delta - 1/16) k$, where $\Delta$ is a divergent, $\vec{k}$-independent constant. 
This divergence should not appear in any physical quantity, and we expected that to be the case once the functional integration is formulated in the global dS space, as discussed in Sec. \ref{sec:discussion}. Since counter terms can always be added to the functional integration measure in \eqref{eq:4d_inner_product} to absorb the divergence, we neglect the latter. Then we find the finite expression
\begin{equation}
    \mathcal{T}^{ab}(\vec{k}) = -\frac{1}{16}k \left(\delta^{ab} + \frac{k^a k^b}{k^2}\right).
\end{equation}

\end{appendices}

\bibliography{bibliography}

\end{document}